\documentclass[preprint,3p,12pt]{elsarticle}

\usepackage{lineno,hyperref}
\modulolinenumbers[5]

\usepackage{hyperref}

\usepackage{epstopdf}

\usepackage{mathrsfs}
\usepackage{amssymb}
\usepackage{amsfonts}
\usepackage{latexsym}
\usepackage{amsmath}
\usepackage{xcolor}
\usepackage{wrapfig}
\usepackage{floatflt}

\usepackage{graphicx}
\def\lb{\label}

\newcommand{\er}[1]{\textrm{(\ref{#1})}}

\newtheorem{theorem}{\bf Theorem}[section]

    \def\cU{{\mathcal U}}

\def\ve{\varepsilon}

    \def\N{{\mathbb N}}   

\def\lt{\biggl}                  \def\rt{\biggr}
               \def\wt{\widetilde}

\let\ge\geqslant                 \let\le\leqslant

\def\iy{\infty}

\def\el2{\ell^{\,2}}             \def\1{1\!\!1}

\def\BBox{\hspace{1mm}\vrule height6pt width5.5pt depth0pt \hspace{6pt}}

\let\ge\geqslant
\let\le\leqslant

\newcommand{\ca}{\begin{cases}}
\newcommand{\ac}{\end{cases}}
\newcommand{\ma}{\begin{pmatrix}}
\newcommand{\am}{\end{pmatrix}}
\def\eq{\begin{equation}}
\def\qe{\end{equation}}
\def\[{\begin{equation}}
\def\]{\end{equation}}

\def\BBox{\hspace{1mm}\vrule height6pt width5.5pt depth0pt \hspace{6pt}}

\begin{document}

\begin{frontmatter}

\title{Fractal mountains and binary random loops}
\date{\today}

\author
{Anton A. Kutsenko}

\address{Jacobs University (International University Bremen), 28759 Bremen, Germany; email: akucenko@gmail.com}

\begin{abstract}
We discuss a variation of Takagi curves based, however, more on algebraic than geometric principles. Namely, we construct functions of loops in a special binary representation. The graph of these functions usually has chaotic and fractal forms, sometimes recall mountain landscapes. Nevertheless, the values in a dense set of points and even the integral of these functions can be calculated explicitly.  
\end{abstract}



\end{frontmatter}


{\section{Introduction}\lb{sec1}}

The first motivation was an attempt to present loops in random walks globally. There are many impressive presentations of particular or a couple of random walks, where steps, turns, returns, or loops are well illustrated. The idea of the current paper is to associate a real number or vector with each infinite walk. After that, we compute the loops multiplied by some weights at each real number. Thus, we obtain some functions that can be analyzed. While such functions have a fractal or chaotic structure, some of their parameters such as $L^1$ and $L^{\iy}$ norms can be evaluated analytically. 

The idea of using the binary representation of the numbers to define some functions goes back to Cantor, Weierstrass, or Leibniz, or even earlier. The resulting functions are very different from the standard smooth ones.
The closest to the functions considered in this work are the so-called Takagi functions or their generalizations: Takagi – Landsberg and de Rham curves, see \cite{T} and recent results in \cite{MS}. The difference is that we use binary expansions based on $+1$, $-1$ instead of $0$, $1$ and we count specifically loops. Thus, our functions are discontinuous and, being deterministic, look stochastic. Also, we make our definitions compatible with self-avoiding walks (SAW). In the sense that if $x$ corresponds to a walk without loops then $U(x)=0$, otherwise $U(x)>0$. This is motivated by the existence of open problems related to the distribution of SAW in the multidimensional case, see \cite{S}.

Despite all this, the objectives of this article are more expository. Namely, we prove some general results and use it to build graphs of functions for special weights of loops and estimate their parameters analytically. Many of the graphs look stochastic, but some resemble mountain landscapes. It is noteworthy that the integrals and some values of such functions can be expressed in terms of well-known constants such as $\pi$, $\ln2$, etc. In somewhat, this article is a continuation of the analysis of special binary representations. Namely, in \cite{K1} special nuber systems are used to show the isomorphism between the algebras of one-dimensional and multidimensional finite-difference operators, the explicit isomorphism has a fractal nature; in \cite{K2} binary representations are used to express all the zeros of Schr\"oder-Poincar\'e entire functions by infinite products of Vi\`ete type, such zeros form fractal structures similar to Julia sets growing up to infinity.

Finally, let us mention other very interesting approaches related to the representation of real numbers as random walks, see \cite{ABBB}.

{\section{Functions that count loops}\lb{sec2}}

For allmost all $x\in[-1,1]$, except a countable set of some dyadic rationals, we can define uniquely their binary representation
\[\lb{e101}
 x=\sum_{n=0}^{+\infty}\frac{x_n}{2^{n+1}},\ \ x_n\in\{-1,+1\}.
\]
For integer $0\le n\le m$, let us define "loop-counting" functions
\[\lb{e102}
 L_{nm}(x)=\ca 1, & \sum_{j=n}^m x_j=0,\\ 0, & {\rm otherwise}. \ac
\]
These functions are well-defined for all $x\in[0,1]$ except a countable set mentioned above. Even if $x$ has different binary expansions, below we sometimes write $L_{nm}(x)$ in view of its concrete given representation. Let $\cU=\{u_{nm}\}_{0\le n\le m<+\iy}$ be some set of non-negative real numbers, $u_{nm}\ge0$. Define the function
\[\lb{e103}
 U(x)=\sum_{0\le n\le m<+\iy}u_{nm}L_{nm}(x).
\]
Typically, this class of functions  has a chaotic or fractal structure. 
\begin{theorem}\lb{T1}
The function $U$ is non-negative, measurable and even. The following identity holds
\[\lb{T11eq}
 \|U\|_{1}=\int_{-1}^{1}U(x)dx=2\sum_{k=1}^{+\iy}\sum_{m,n\ge0,\ m-n=2k-1}4^{-k}\binom{2k}{k}u_{nm}.
\]	
The essential supremum of $U$ is reached at $x=\pm\frac13$ and equal to
\[\lb{T12eq}
 \|U\|_{\iy}=U\lt(\frac{\pm1}3\rt)=\sum_{k=1}^{+\iy}\sum_{m,n\ge0,\ m-n=2k-1}u_{nm}.
\] 
\end{theorem}
{\it Proof.} Consider a finite dyadic rational number
$$
 \wt x=\sum_{n=0}^m\frac{\wt x_n}{2^{n+1}}
$$
for some $\wt x_n\in\{-1,+1\}$ and $m\ge0$. Almost all numbers $x\in[\wt x-\ve,\wt x+\ve]$ for $\ve=2^{-m-1}$ has the same initial segment in its binary representation as $\wt x$, i.e. $x_n=\wt x_n$, $0\le n\le m$. Moreover, if $x\not\in[\wt x-\ve,\wt x+\ve]$ then $x_n\not=\wt x_n$ for some $0\le n\le m$. Thus, the piece-wise constant functions
$$
 U_m(x)=\sum_{\wt x_0,...,\wt x_m\in\{-1,+1\}}\chi_{[\wt x-2^{-m-1},\wt x+2^{-m-1}]}(x)\sum_{n=0}^mu_{nm}L_{nm}(\wt x)
$$
converge from below $U_m(x)\nearrow U(x)$, $m\to+\iy$, since the corresponding characteristic functions cover the whole interval
$$
 \sum_{\wt x_0,...,\wt x_m\in\{-1,+1\}}\chi_{[\wt x-2^{-m-1},\wt x+2^{-m-1}]}(x)=\chi_{[-1,1]}(x)\ \ {\rm upto\ a\ zero\ Lebesgue\ measure.}
$$
Hence, $U(x)$ is measurable and by the monotone convergence Theorem we have
\begin{multline}
 \int_{-1}^1U(x)dx=\lim_{m\to+\iy}\int_{-1}^1U_m(x)dx=\lim_{m\to+\iy}\sum_{\wt x_0,...,\wt x_m\in\{-1,+1\}}2^{-m}\sum_{n=0}^mu_{nm}L_{nm}(\wt x)\\
 =\lim_{m\to+\iy}2^{-m}\sum_{m-n+1\in2\N\cap\{0,...,m\}}u_{nm}2^{n}\binom{m-n+1}{\frac{m-n+1}2}=2\sum_{k=1}^{+\iy}\sum_{m,n\ge0,\ m-n=2k-1}4^{-k}\binom{2k}{k}u_{nm},\notag
\end{multline}
since $L_{nm}(\wt x)=1$ if and only if the quantities of $-1$ and $+1$ among $\{\wt x_n,...,\wt x_m\}$ are equal to each other, otherwise $L_{nm}(\wt x)=0$. Such quantities coincide with the binomial coefficients $\binom{m-n+1}{\frac{m-n+1}2}$ if $m-n$ is odd and they are $0$ if $m-n$ is even. Now, using the fact that the first components $\wt x_0,...,\wt x_{n-1}$ do not affect on $L_{nm}(\wt x)$ we obtain that the number of $\wt x_0,...,\wt x_m$ for which $L_{nm}=1$ is exactly $2^n\binom{m-n+1}{\frac{m-n+1}2}$ for odd $m-n$ and $0$ otherwise.

The function $U(x)$ reaches its maximal value when the number of loops is maximal too. This happens for $x$ having alternating signs in the binary representation, i.e. $x_0=+1$, $x_1=-1$, $x_2=+1$ and so on or $x_0=-1$, $x_1=+1$, $x_2=-1$ and so on. Hence, $x=\pm\frac13$. The values $U(\pm\frac13)$ can be evaluated explicitly by using \er{e102}, \er{e103}, they coincide with \er{T12eq}.

The fact that $U$ is even function is obvious, since $L_{nm}$ are even. \BBox

The results of Theorem \ref{T1} can be directly extended to the multidimensional case
\[\lb{e104}
 U(x,...,w)=\sum_{0\le n\le m<+\iy}u_{nm}L_{nm}(x)...L_{nm}(w),\ \ \ (x,...,w)\in[-1,1]^N.
\]
\begin{theorem}\lb{T2}
The function $U$ is non-negative, measurable, and even by each of its argument. The following identity holds
\[\lb{T21eq}
 \|U\|_{1}=\int_{-1}^{1}...\int_{-1}^{1}U(x,...,w)dx...dw=2\sum_{k=1}^{+\iy}\sum_{m,n\ge0,\ m-n=2k-1}4^{-Nk}\binom{2k}{k}^Nu_{nm}.
\]	
The essential supremum of $U$ is reached at $x=\pm\frac13$,...,$w=\pm\frac13$ and equal to
\[\lb{T22eq}
 \|U\|_{\iy}=U\lt(\frac{\pm1}3,...,\frac{\pm1}3\rt)=\sum_{k=1}^{+\iy}\sum_{m,n\ge0,\ m-n=2k-1}u_{nm}.
\] 
\end{theorem}

{\bf Remark.} For $u_{nm}>0$, definition \er{e104} is compatible with the definition of self-avoiding random walk. If $(x,...,w)$ corresponds to the walk without loops, then $U(x,...,w)=0$, otherwise $U(x,...,w)>0$. The coefficients $u_{nm}$ represent the weights of loops. It makes sense to assume that the loops that appear far from the starting point have less weight. The assumption is necessary for bounding $\|U\|_{1}$ or/and $\|U\|_{\iy}$.

{\section{Examples}\lb{sec3}}

We focus on the case when $u_{nm}$ do not depend on $n$. If $u_{nm}$ are monotonic then the convergence condition for the series \er{T11eq}, \er{T12eq} and \er{T21eq}, \er{T22eq} can be obtained from the asymptotics
$$
 4^{-k}\binom{2k}{k}\sim\frac1{\sqrt{\pi k}}.
$$ 
But in many of the examples presented, the corresponding series can be computed explicitly.

{\bf 1.} We consider $\cU$ consisting of
\[\lb{200}
 u_{nm}=\frac{4}{(m+2)(m+3)(m+4)}.
\]
In this case we can compute uniform and integral norms explicitly. Since 
$U$ is even, we focus on the interval $[0,1]$. Due to Theorem \ref{T1}, the $L^1$-norm is
\begin{multline}
 \int_0^1U(x)dx=\sum_{k=1}^{+\iy}\sum_{m,n\ge0,\ m-n=2k-1}4^{-k}\binom{2k}{k}\frac{4}{(m+2)(m+3)(m+4)}=\\
 \sum_{k=1}^{+\iy}4^{-k}\binom{2k}{k}\sum_{m=2k-1}^{+\iy}\frac{4}{(m+2)(m+3)(m+4)}=\sum_{k=1}^{+\iy}4^{-k}\binom{2k}{k}\frac{2}{(2k+1)(2k+2)}=\\
 2\int_0^1(\arcsin x-x)dx=\pi-3.\notag
\end{multline}
In turn, the $L^{\iy}$-norm is 
\begin{multline}
 U\lt(\frac13\rt)=\sum_{k=1}^{+\iy}\sum_{m,n\ge0,\ m-n=2k-1}\frac{4}{(m+2)(m+3)(m+4)}=
 \sum_{k=1}^{+\iy}\sum_{m=2k-1}^{+\iy}\frac{4}{(m+2)(m+3)(m+4)}=\\ 
 \sum_{k=1}^{+\iy}\frac{2}{(2k+1)(2k+2)}=2\sum_{k=1}^{+\iy}\lt(\frac{1}{2k+1}-\frac1{2k+2}\rt)=2\sum_{k=3}^{+\iy}\frac{(-1)^{k+1}}{k}=2\ln2-1.
 \notag
\end{multline}
It is also possible to compute $\underline{\lim}_{x\to0}U(x)=\frac1{15}$ by using the binary expansion 
$$
 0=[+1,-1,-1,-1,-1,-1,-1,-1,...]
$$ 
having one loop only. The next discussion is rough but it reflects the structure of $U$ in some sense. Dyadic rationals $x=\frac{p}{2^n}$, $p\in\N$ have a finite number of loops. They corresponds to local minimums, and $U(x)$ is a rational number for such $x$. The rationals $x=\frac{p}{2^n3}$, $p\not\in3\N$ have a maximal number of loops, since their tails are $[...,-1,+1,-1,+1,-1,+1,...]$. The corresponding values $U(x)$ are local maximums, they are transcendental numbers that can be expressed in terms of $\ln2$ and some rational numbers.
\begin{figure}
  \centering
    \includegraphics[width=0.99\textwidth]{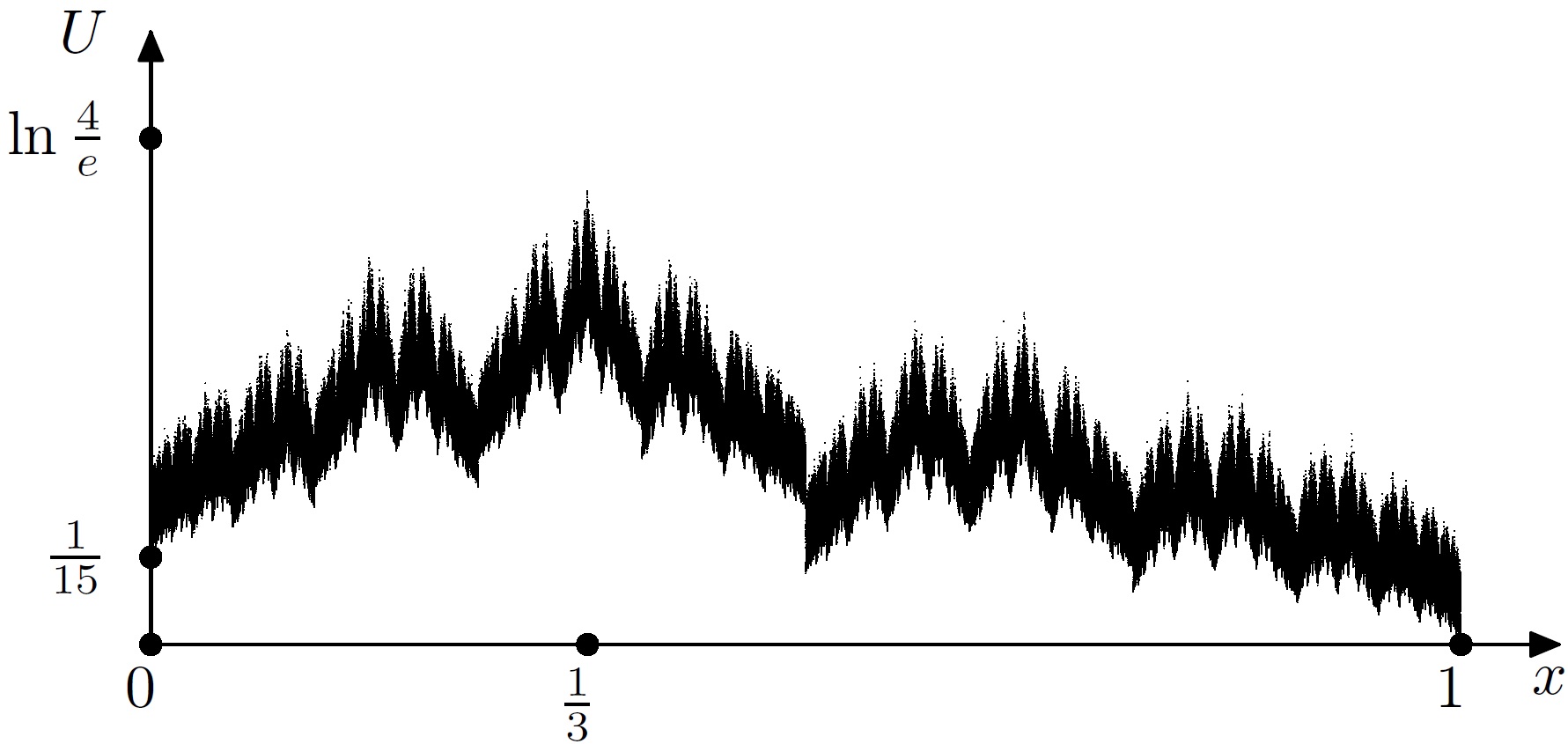}
  \caption{A random generator of $10^6$ binary samples of the length $\approx100$ symbols $\pm1$ is used to plot the graph of $U$ corresponding to the weights \er{200}.\lb{fig1}} 
\end{figure}
The graph of $U$ is plotted in Fig. \ref{fig1}. The same selective random generator was applied for both: plots and numerical integrations. In particular, the Monte Carlo method along with the generator give $\approx0.141$ for $\int_0^1U(x)dx$ that is close to the analytic result $\pi-3$. 

{\bf 2.} Now, let us consider $\cU$ consisting of
\[\lb{201}
 u_{nm}=\frac{1}{(m+2)^2}-\frac{1}{(m+3)^2}.
\]
The weights are similar to \er{200} but the computation of $L^1$-norm is a little bit more complex
\begin{multline}
 \int_0^1U(x)dx=\sum_{k=1}^{+\iy}\sum_{m,n\ge0,\ m-n=2k-1}4^{-k}\binom{2k}{k}\lt(\frac{1}{(m+2)^2}-\frac{1}{(m+3)^2}\rt)=\\
 \sum_{k=1}^{+\iy}4^{-k}\binom{2k}{k}\sum_{m=2k-1}^{+\iy}\lt(\frac{1}{(m+2)^2}-\frac{1}{(m+3)^2}\rt)=\sum_{k=1}^{+\iy}4^{-k}\binom{2k}{k}\frac{1}{(2k+1)^2}=\\
 \int_0^1\frac{\arcsin x-x}xdx=\int_0^{\frac\pi2}x\cot xdx-1=-\int_0^{\frac\pi2}\ln\sin xdx-1=\frac{\pi\ln2-2}2,\notag
\end{multline}
since it is known that
\begin{multline}
 \int_0^{\frac\pi2}\ln\sin xdx=\frac{\int_0^{\frac\pi2}\ln\sin xdx+\int_0^{\frac\pi2}\ln\cos xdx}2=\frac{\int_0^{\frac\pi2}(\ln\sin 2x-\ln2)dx}2=\\ \frac14\int_0^{\pi}\ln\sin xdx-\frac{\pi\ln2}4=\frac12\int_0^{\frac\pi2}\ln\sin xdx-\frac{\pi\ln2}4\ \Rightarrow\ \int_0^{\frac\pi2}\ln\sin xdx=-\frac{\pi\ln2}2.\notag
\end{multline}
The $L^{\iy}$-norm is 
\begin{multline}
 U\lt(\frac13\rt)=\sum_{k=1}^{+\iy}\sum_{m,n\ge0,\ m-n=2k-1}\lt(\frac{1}{(m+2)^2}-\frac{1}{(m+3)^2}\rt)=\\
 \sum_{k=1}^{+\iy}\sum_{m=2k-1}^{+\iy}\lt(\frac{1}{(m+2)^2}-\frac{1}{(m+3)^2}\rt)= 
 \sum_{k=1}^{+\iy}\frac{1}{(2k+1)^2}=\frac{\pi^2}8-1.
 \notag
\end{multline}
\begin{figure}
  \centering
    \includegraphics[width=0.99\textwidth]{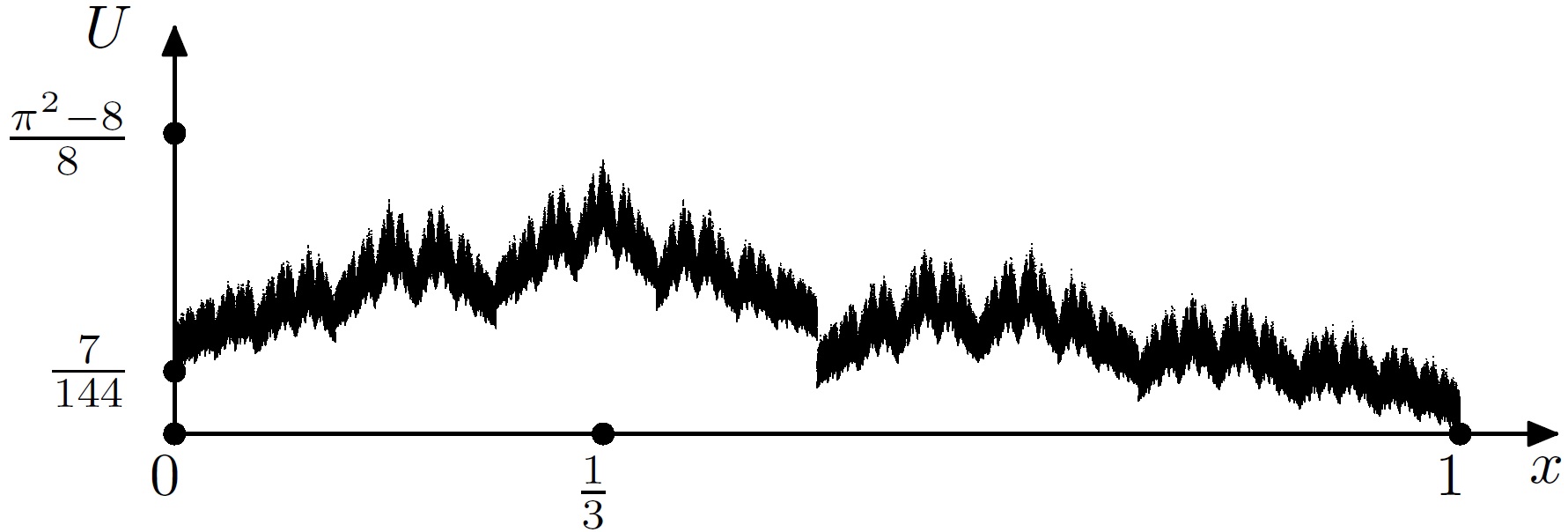}
  \caption{A random generator of $10^6$ binary samples of the length $\approx100$ symbols $\pm1$ is used to plot the graph of $U$ corresponding to the weights \er{201}. \lb{fig2}}  
\end{figure}
The graph of $U$ is plotted in Fig. \ref{fig2}.

{\bf 3.} Now, let us consider $\cU$ consisting of
\[\lb{202}
 u_{nm}=\frac{1}{(m+0.7)^{3.25}}.
\]
The computation of the maximum and integral perhaps cannot be expressed in terms of standard values as $e$, $\pi$, etc., but it may include some values of Hurwitz zeta function. We simply plot the graph of $U$ which may recall mountains on some Chinese and Japanese landscapes, see Fig. \ref{fig3}.
\begin{figure}
  \centering
    \includegraphics[width=0.99\textwidth]{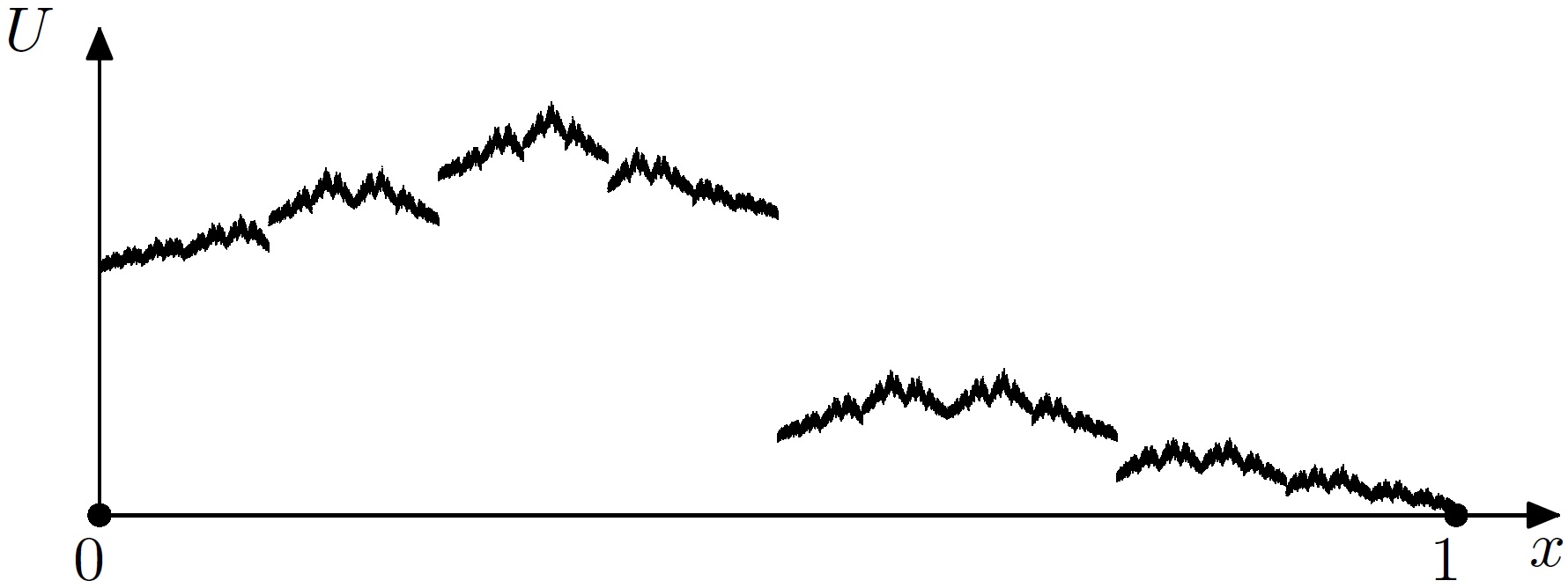}
  \caption{A random generator of $10^6$ binary samples of the length $\approx100$ symbols $\pm1$ is used to plot the graph of $U$ corresponding to the weights \er{202}. \lb{fig3}}  
\end{figure}

{\bf 4.} Finally, we consider $\cU$ consisting of
\[\lb{203}
u_{nm}=2^{-m-1}.
\]
The $L^1$-norm is
\begin{multline}
	\int_0^1U(x)dx=\sum_{k=1}^{+\iy}\sum_{m,n\ge0,\ m-n=2k-1}4^{-k}\binom{2k}{k}2^{-m-1}=\sum_{k=1}^{+\iy}4^{-k}\binom{2k}{k}\sum_{m=2k-1}^{+\iy}2^{-m-1}=\\
	\sum_{k=1}^{+\iy}4^{-k}\binom{2k}{k}2^{-2k+1}=\frac{2}{\sqrt{1-2^{-2}}}-2=\frac4{\sqrt{3}}-2.\notag
\end{multline}
The $L^{\iy}$-norm is 
$$
	U\lt(\frac13\rt)=\sum_{k=1}^{+\iy}\sum_{m,n\ge0,\ m-n=2k-1}2^{-m-1}=
	\sum_{k=1}^{+\iy}\sum_{m=2k-1}^{+\iy}2^{-m-1}=\sum_{k=1}^{+\iy}2^{-2k+1}=\frac23.
$$
\begin{figure}
	\centering
	\includegraphics[width=0.9\textwidth]{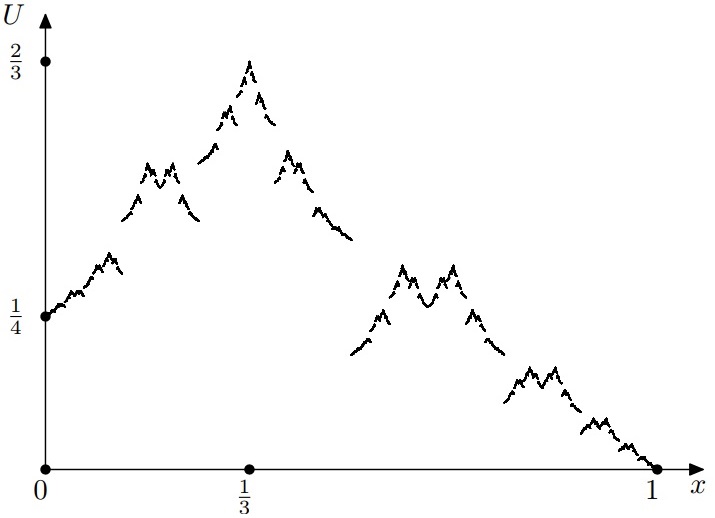}
	\caption{A random generator of $10^6$ binary samples of the length $\approx100$ symbols $\pm1$ is used to plot the graph of $U$ corresponding to the weights \er{203}. \lb{fig4}}  
\end{figure}
While $U$ is discontinuous, its plot, see Fig. \ref{fig4}, looks "smoother" than other plots in this Section, because the scales of variance in both coordinates are the same $2^{-n}$.

{\bf 5.} One more example relates to the two-dimensional $U(x,y)$ with the weigths \er{203}. This example requires much more computations than one-dimensional ones. We provide the sketch only, see Fig. \ref{fig5}. Due to Theorem \ref{T2}, the maximum of $U$ is again $U(\frac13,\frac13)=\frac23$.

\begin{figure}
	\centering
	\includegraphics[width=0.5\textwidth]{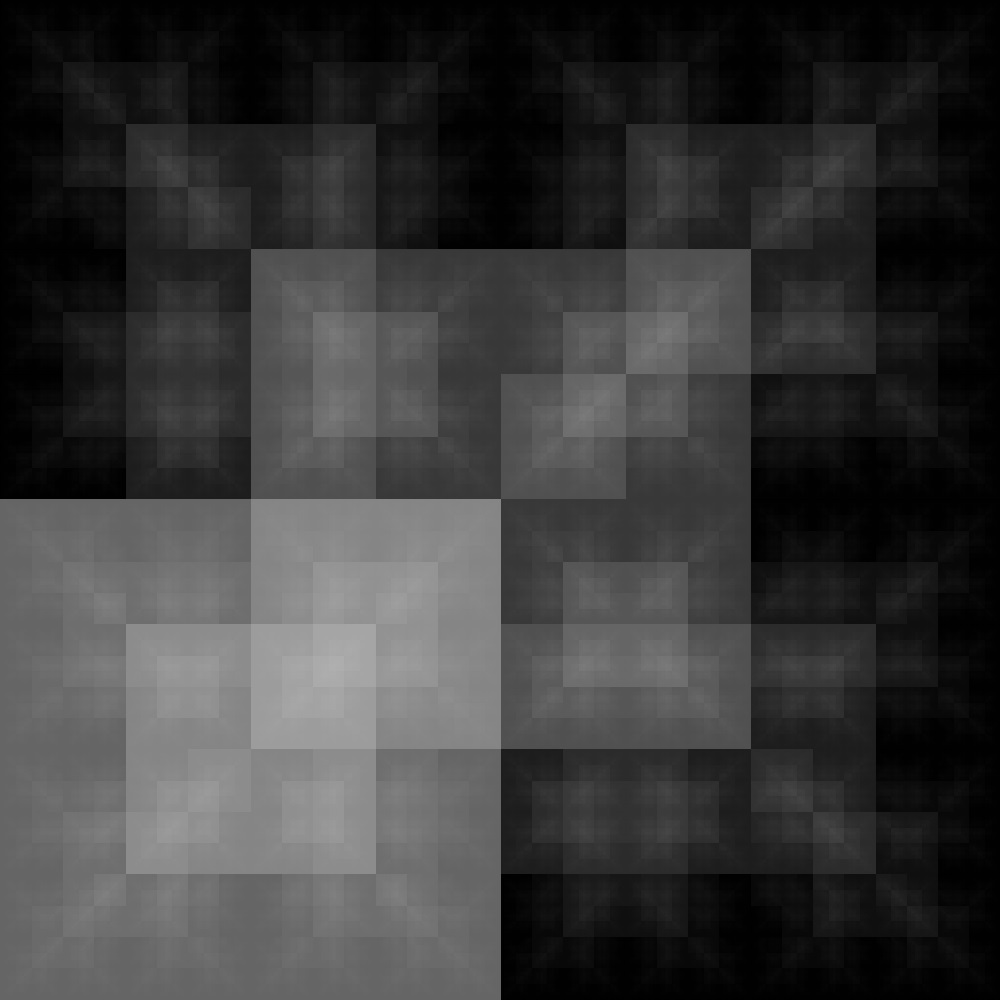}
	\caption{Sketch of $U(x,y)$, $(x,y)\in[0,1]^2$ with the weights \er{203}. Black points correspond to small values of $U$, white poits mean relatively large values. \lb{fig5}}  
\end{figure}

\section*{Funding statement}
This paper is a contribution to the project M3 of the Collaborative Research Centre TRR 181 "Energy Transfer in Atmosphere and Ocean" funded by the Deutsche Forschungsgemeinschaft (DFG, German Research Foundation) - Projektnummer 274762653. This work is also supported by the RFBR (RFFI) grant No. 19-01-00094.

\end{document}